# From Sedimentation to Suspension: Critical Strain as a Predictor of Particle Resuspension Thresholds

Mohammadreza Mahmoudian,[1] Simon A. Rogers,[2] and Parisa Mirbod[1]

[1] *Department of Mechanical and Industrial Engineering, University of Illinois Chicago, Chicago, IL 60607, USA*

[2] *Department of Chemical and Biomolecular Engineering, University of Illinois at Urbana-Champaign, Urbana, IL 61801, USA*

(*Electronic mail: pmirbod@uic.edu.)

(Dated: 14 November 2025)

Viscous resuspension, the process by which sedimented particles are re-entrained into a fluid under flow, is central to numerous natural and industrial systems, including environmental contaminant transport, riverbed erosion, and biogeochemical cycling. Despite its ubiquity and importance, predicting when and how resuspension occurs remains challenging, particularly under oscillatory shear, where particle interactions are nonlinear, collective, and time-dependent. Here, we examine the resuspension dynamics of dense, non-Brownian suspensions under both steady and oscillatory shear using bulk rheometry and in situ rheo-microscopy over a broad range of particle volume fractions ($\phi$= 0.30 to 0.55). We demonstrate that strain is the key control parameter governing the transition from a sedimented bed to a fully suspended state. This strain-driven onset is mediated by effective interparticle collisions and collective particle motion. We develop a predictive model that captures the observed strain thresholds as a function of volume fraction, allowing for the construction of a new state diagram delineating sedimentation, resuspension, and full suspension regimes. These findings reveal a robust, strain-controlled resuspension mechanism and establish a unified framework for predicting suspension behavior across steady and oscillatory flows, offering new tools for managing particle-laden transport in geophysical, biological, and industrial environments.

## I. INTRODUCTION

The resuspension of particles that have previously settled in Newtonian fluids is important in both natural and industrial environments. In suspensions in which the particles are denser than the surrounding fluid, known as negatively buoyant suspensions, gravity causes the particles to sediment over time. However, under certain flow conditions, hydrodynamic forces and particle interactions can counteract sedimentation, leading to a reversal of this process, known as viscous resuspension[1–3]. This phenomenon has been observed across a wide range of temporal and spatial scales, from milliseconds and micrometers in microfluidic devices to decades and kilometers in sediment transport through rivers and coastal environments.

Understanding the mechanisms of viscous resuspension is crucial, as it has far-reaching implications for both environmental health and industrial efficiency. In natural environments, the resuspension of particles in aquatic systems drives essential processes such as sediment transport, erosion, and pollutant distribution, which directly influence the physical landscape and the health of these ecosystems. Resuspension plays a significant role in the dispersal of sediments, microplastics, and pollutants in oceans, lakes, and rivers[4–9]. The presence of resuspended sediments can increase turbidity levels, impairing light penetration, thereby affecting photosynthetic organisms. Further, sediments can act as carriers for pollutants, posing health risks to wildlife and humans. Accurately predicting the dynamics of resuspension is therefore essential for effective water resource management, ecosystem conservation, and adaptation to challenges posed by environmental changes.

In engineered systems, controlling viscous resuspension is equally critical. Industries such as composite processing, mining, and wastewater treatment rely on the precise manipulation of particle-laden flows to optimize processes, ensure product quality, and mitigate environmental impacts. For example, in chemical production, maintaining uniform particle distributions within fluids is vital for ensuring consistent reactions and adhering to product standards. In mining operations, managing sediment resuspension is crucial for minimizing material losses and preventing contamination[10,11]. In wastewater treatment, controlling resuspension facilitates pollutant removal and supports overall system functionality. Failure to manage resuspension can lead to significant inefficiencies, increased costs, and environmental hazards. Although resuspension has a negative impact in some cases, in other scenarios, it is crucial to optimize resuspension characteristics. For example, in ultrasonic cleaning, resuspension aids in the detachment and removal of deposits from surfaces, increasing cleaning efficacy[12]. This dual nature of resuspension, as both a potential hazard and a beneficial process, highlights the need for a comprehensive understanding of its mechanisms across different timescales, under dynamic conditions and steady flows, and in various environments.

In general, sediment transport begins when the viscous forces from fluid flow over a sediment bed exceed the gravitational forces acting on the particles. This balance of forces is typically expressed by the Shields number, $\tau^* = \tau/(\Delta\rho g d)$, where $\tau$ is the shear stress, $\Delta\rho$ is the density difference between the particles and the fluid, $g$ is the acceleration due to gravity, and $d$ represents the particle diameter[6,13,14]. When the shear stress exceeds the critical Shields number ($\tau^* = 1$), bed load transport begins. During bed load transport, a thin layer of particles on the surface moves while maintaining frequent contact with the static granular bed below[15–18].

As the shear stress increases further, the particles become resuspended, creating a vertical volume fraction gradient that eventually dissipates, leading to a uniform suspension[1–3].

The study of viscous resuspension in laboratory settings typically involves carefully designed rheological experiments in which shear stresses and strain rates can be accurately measured. In their foundational work, Leighton and Acrivos first established a quantitative relationship between the steady-state resuspension height and the Shields number, balancing the effects of shear-induced diffusion against gravitational settling[2,3]. Building on their pioneering work, extensive numerical and experimental research has been performed to further explore the parameters affecting resuspension dynamics[19–22], including particle density, size distribution (polydispersity), and shape[23–27]. These factors are now recognized as significant determinants of resuspension behavior across different environmental and industrial contexts.

While the Shields number remains a central metric for predicting particle resuspension, its application in turbulent flows is limited due to spatial and temporal fluctuations in shear stress that complicate threshold-based predictions[28–31]. These transient forces can intermittently destabilize sediment beds, making resuspension a dynamic, rather than static, process. In parallel, particle-particle interactions, such as frictional contacts and clustering, play a critical but not fully understood role in shaping resuspension behavior and sediment transport patterns[32–35]. These interactions affect both the microscale rheology and the macroscale evolution of sediment beds. Recent studies under laminar shear conditions have further challenged traditional threshold-based models, revealing a continuous transition from creeping motion to granular flow, rather than a discrete onset[18,36–38]. This transition is governed by the viscous number, a dimensionless measure of the balance between viscous and particle interaction forces, and occurs at a fixed concentration, suggesting a rheological rather than purely force-driven mechanism. Together, these findings call for new frameworks that account for the role of flow history, microstructure, and shear mode in governing resuspension.

Despite significant progress in understanding viscous resuspension in turbulent and inertia-dominated flows, far less is known about systems where gravitational settling plays a comparably significant role. In particular, resuspension dynamics under laminar shear, where the Shields number is approximately 1, has received limited attention. Unlike turbulent regimes where particle motion is dominated by inertial bursts and chaotic eddies, these mixed-force environments are governed by a delicate balance between shear-induced lift and gravity-driven settling. Such conditions are frequently encountered in many natural and engineered systems that operate at low Reynolds numbers. Yet the fundamental physical mechanisms that dictate resuspension in these laminar, gravity-influenced flows remain poorly understood, motivating the need for a detailed investigation of how shear, particle interactions, and microstructure interplay to drive suspension dynamics.

Several important examples highlight the ubiquity and impact of laminar oscillatory flow on resuspension phenomena. In the cardiovascular system, microparticle transport and deposition occur under laminar, pulsatile blood flow, influencing processes such as thrombosis, embolism formation, and targeted drug delivery. In riverbeds, estuaries and coastal environments, oscillatory laminar flows induced by tides or wave action control the mobilization and deposition of sediments. Similarly, during seismic events, transient laminar stress fields can reconfigure subsurface sediments, triggering liquefaction, landslides, or the release of buried contaminants. Unlike turbulence-driven resuspension, where mixing is enhanced by chaotic fluctuations, laminar resuspension involves distinct thresholds, intermittency, and memory effects that can dramatically influence system behavior and stability.

To address these critical gaps, we investigate the resuspension dynamics of dense, non-Brownian suspensions in Newtonian fluids under both steady and oscillatory laminar shear flow. Using a combination of bulk rheological measurements and in situ rheo-microscopy across a broad range of particle volume fractions, we show that a strain-driven mechanism governs the transition from a sedimented bed to a dynamically suspended particle state. Our results demonstrate that shear strain, rather than shear rate, is the dominant control parameter for resuspension in both steady and oscillatory flow regimes. That is, while the steady-state studies have demonstrated the importance of shear rates in maintaining a spatial concentration gradient, we show that strain-dependent phenomena are required to establish steady-state conditions. To support these experimental findings, we develop a semi-empirical predictive framework that quantitatively links the critical strain for resuspension to the particle volume fraction. Once calibrated from a single dataset, this framework predicts the critical strain thresholds across different concentrations and shear modes By integrating these findings, we construct state diagrams that delineate sedimentation, resuspension, and suspension states. This framework provides a predictive understanding of particle transport under mixed gravitational and shear forces. Beyond advancing the fundamental physics of laminar resuspension, our model has broad implications, including providing a way to develop informed strategies to minimize unwanted particle mobilization in natural ecosystems, to improve the efficiency of industrial processes involving particulate flows, and to support environmental resilience and biomedical innovation.

## II. MATERIALS AND METHODS

### A. Materials and sample preparation

The polyethylene spheres used in this investigation were obtained from Cospheric. The spheres had diameters ranging from 75–90 $\mu$m and a density of 1.25 $g/cm^3$. These particles were dispersed in AR20 silicone oil ($\rho$= 1.00 $g/cm^3$) purchased from Sigma-Aldrich, which behaves as a Newtonian fluid with a viscosity of $\eta_m$= 0.02 Pa·s. A gravitational force is generated owing to the density difference between the particles and the silicone oil, ensuring sedimentation in the suspension system. To prepare the suspension, the particles were dispersed in the Newtonian silicone oil using vortex mixing for 60 s. Suspensions with particle volume fractions of 0.30, 0.40, 0.50, and 0.55 were made. These concentrations were selected to explore a range of dense particle suspensions. The particle volume fraction in the suspension, $\phi$, was calculated using the following equation[39]:

$$\phi_m = \frac{\rho_m x}{\rho_m x + \rho_d (1-x)}, \quad (1)$$

where $\rho_m$, $\rho_d$, and x are the liquid density, particle density, and mass fraction of the particles, respectively.

### B. Rheological measurements

Rheological measurements were conducted using a DHR-2, a rotational rheometer from TA Instruments with a parallel plate geometry that has a 50 mm diameter. The gap size between the two parallel plates was kept at 1 mm. The temperature was maintained at 20 °C, at which the viscosity of the silicone oil was approximately $\eta_m$= 0.02 Pa·s. To eliminate slip, sandpaper was applied to both the upper and lower plates, although no significant differences were observed between measurements with smooth and rough surfaces. Before the rheological tests, all the samples were presheared for 120 s at a shear stress of $\tau$ = 10 Pa, which is sufficient to ensure a fully developed and fully suspended steady-state flow. The samples were allowed to rest under zero shear stress for various periods to enable sedimentation before additional experimentation. It was visually confirmed that the samples remained intact throughout all reported measurements.

The oscillatory shear measurements were conducted using a data acquisition setting with zero seconds of conditioning time and three cycles per data point. Additional details regarding the influence of the number of cycles on the resuspension behavior are provided in the Supplementary Information.

### C. Optical imaging and microstructure visualization

To capture the flow field and particle dynamics within the rheometer gap, a high-speed camera (MIRO LAB 320 from Phantom) was employed. This imaging system is equipped with a 2X F-mount adapter tube and an adjustable 12X zoom lens provided by Navitar. The setup offers a sufficient aperture to record images with a resolution of 1920 by 1200 pixels, ensuring clear visualization of the processes occurring within the suspension. Additionally, a 150 W standalone LED (MultiLed QT from GSVITEC) was used as the light source, providing optimal illumination for high-speed imaging.

## III. RESULTS

### A. Strain-Governed Particle Resuspension and Scaling Mechanisms under Oscillatory Shear Flow

To uncover the physical mechanisms governing shear-induced resuspension in dense, non-Brownian suspensions, we conducted a series of oscillatory rheometry experiments, systematically varying strain amplitude, volume fraction, and sedimentation time. Our goal was to determine how deformation, rather than force balance, governs the transition from a gravitationally sedimented bed to a fully resuspended state.

Experiments were conducted using PE particles dispersed in a Newtonian silicone oil with a slight density mismatch ($\Delta\rho \approx 0.25$ g/cm$^3$), enabling controlled sedimentation, see Methods for more details. Suspensions were vortex-mixed and pre-sheared at 10 Pa to reset microstructural history, then allowed to rest for durations ranging from 0 to 900 seconds. Following rest, sinusoidal shear strain of the form $\hat{\gamma}(t) = \hat{\gamma}_0 \sin(\omega t)$ was applied in a parallel-plate rheometer with simultaneous side-view imaging (Figure 1(a)).

We show in Figure 1(b) that the magnitude of the complex viscosity decreased with small strain amplitudes because the hydrodynamic forcing was insufficient to overcome gravitational sedimentation. As a result, a vertical particle concentration gradient formed. This decrease in the magnitude of the complex viscosity was more pronounced at shorter rest times (e.g., 0-60 s), while at longer rest durations ($\geq$ 300 s), the magnitude of the complex viscosity plateaued at a low value, indicating the establishment of a stable sediment bed.

According to the definition of shear strain in particulate systems, and the critical strain needed to bring neighboring particles into contact[40,41], a critical strain amplitude, $\hat{\gamma}_{col}$, can be identified. Below this threshold, particles indicate minimal rearrangement and a stable sedimented structure, while above it, collisions between particles begin to occur. This threshold is referred to as the collision strain and is defined as:

$$\hat{\gamma}_{col} = \frac{\delta}{2a}, \quad (2)$$

where, $\delta = v^{-1/3} - 2a$ represents the intersurface distance between particles, $v$ is the number density, and $a$ is the particle radius[41]. For a particle volume fraction $\phi$ of 0.50, we estimated $\hat{\gamma}_{col} \approx 0.06$, indicated by the vertical dotted line in Figure 1(b).

Beyond this threshold, the first critical strain amplitude, $\hat{\gamma}_i = 0.38$, was identified at which the viscosity of the complex began to increase due to particles detachment, marking the beginning of active resuspension. At a second critical strain amplitude of $\hat{\gamma}_c = 5.16$, the complex viscosity reached a steady-state plateau, indicating a fully resuspended and fully dispersed particle phase. The transition range, bounded by these critical amplitudes, was consistent across all rest durations, confirming the robustness of the strain-induced resuspension mechanism. Accompanying schematics in Figure 1(b) represent the microstructural changes during this transition. For a more quantitative interpretation of the resuspension dynamics, we performed a regime-based scaling analysis, which is discussed in detail in the Supplementary Note.

Figure 1(c) reinforces the strain-governed resuspension framework through rheological decomposition of the complex viscosity into its viscous ($\eta'$) and elastic ($\eta''$) components. Across the full strain amplitude range ($\hat{\gamma}_0 \approx 0.01$–10), $\eta'$ consistently dominates $\eta''$ by nearly an order of magnitude, with an average $\eta'/\eta''$ ratio of $\sim 10$. This confirms that viscous dissipation, not elastic energy storage, governs the suspension response. This behavior reflects the Newtonian character of the silicone oil medium and the absence of structural elasticity in the non-Brownian, weakly interacting PE particle bed. Unlike viscoelastic or turbulence-driven systems where elasticity or inertia dominates[29,33,42], here the transition from sedimentation to resuspension is driven purely by irreversible shear-induced particle rearrangements. The strong dominance of $\eta'$ and the lack of frequency dependence highlight accumulated strain amplitude as the primary control parameter.

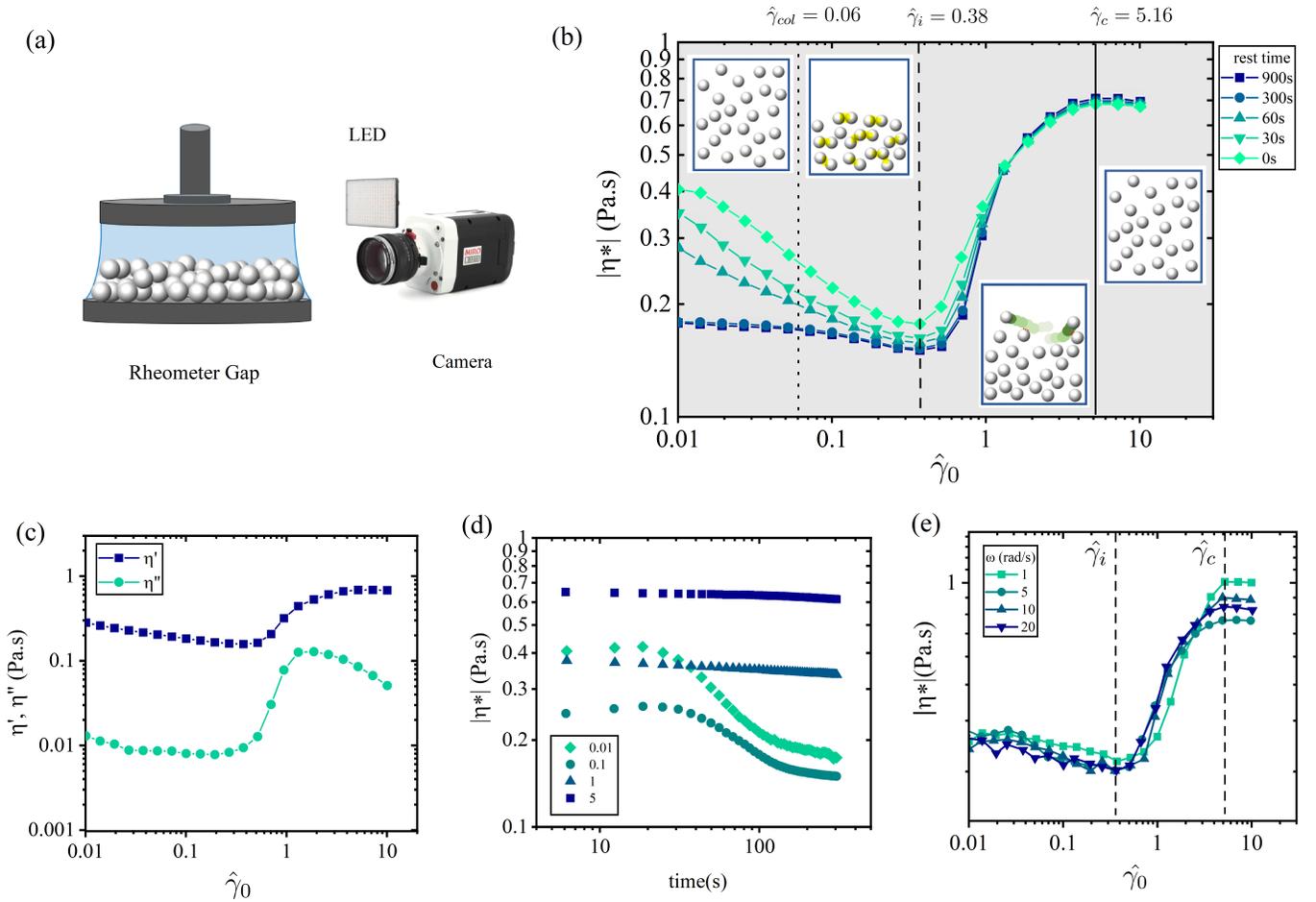

FIG. 1. (a) Schematic of the rheometer setup with side-view imaging for particle visualization. (b) Magnitude of complex viscosity $|\eta^*|$ as a function of strain amplitude for different rest times (0-900 s) at $\phi = 0.50$. The viscosity initially decreases due to sedimentation, then the collision strain $\hat{\gamma}_{col} \approx 0.06$ (dotted line), marking the onset of particle contacts, followed by particle detachment at $\hat{\gamma}_i = 0.38$ (dashed line), where upward motion, rolling, and collision increase viscosity; and a full suspension state is achieved at $\hat{\gamma}_c = 5.16$ (solid line). (c) Real ($\eta'$, elastic) and imaginary ($\eta''$, viscous) components of $\eta^*$ versus strain amplitude at a fixed rest time of 30 s and $\phi = 0.50$. (d) Time evolution of $|\eta^*|$ under continuous oscillatory shear at angular frequency $\omega = 10$ $rad/s$ for strain amplitudes of 0.01, 0.1, 1, and 5. (e) Strain sweep measurements performed under different angular frequencies: 1, 5, 10, and 20 rad/s. The values of the error bars are smaller than the symbol size and are thus not shown here.

We then conducted time-sweep measurements at various strain amplitudes, to test the temporal robustness of these states. Each test was initiated with a preshear at a high stress level ($\tau = 10$ Pa) for 120 s to eliminate flow history effects and ensure uniform dispersion. As shown in Figure 1(d), Oscillatory shear was then immediately applied at fixed amplitudes of 1%, 10%, 100%, and 500%, while monitoring viscosity as a function of time. At low strain amplitudes (1%– 10%), viscosity exhibits a gradual decay, indicative of progressive sedimentation as gravitational settling dominates over shear-induced agitation. This result highlights that without a critical strain input, sedimentation remains the dominant mechanism, and the suspension transitions back toward a phase-separated state. At high amplitudes (100%–500%),the viscosity remains stable over time, indicating that the suspension remains resuspended. These strain levels exceed the previously identified critical strain threshold $\hat{\gamma}_i = 0.38$, corresponding to the onset of resuspension, and thus provide sufficient hydrodynamic agitation to counteract gravitational settling. This sustained response implies that once the applied strain surpasses a critical threshold, particle collisions and upward transport become persistent, maintaining a dynamically stable state.

To evaluate the influence of oscillation frequency on the

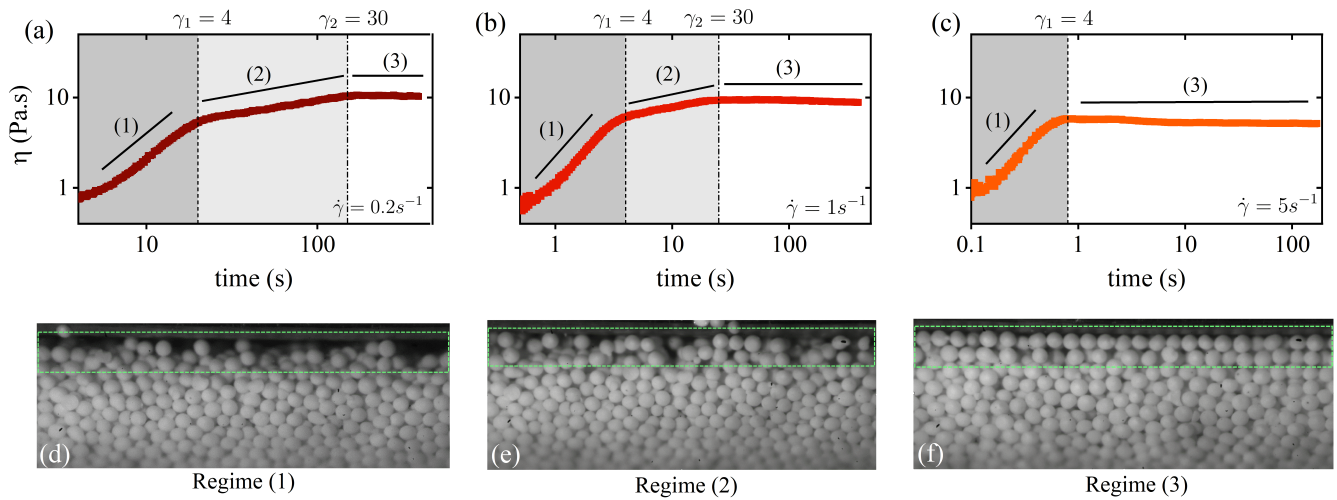

FIG. 2. Transient evolution of viscosity and microstructure during shear startup ($\phi = 0.55$). (a-c) Time-dependent viscosity profiles measured under constant shear rates of (a) $\dot{\gamma} = 0.2\,s^{-1}$, (b) $\dot{\gamma} = 1\,s^{-1}$, and (c) $\dot{\gamma} = 5\,s^{-1}$. The vertical dashed and dashed-dotted lines mark characteristic strain thresholds, $\gamma_1 = 4$ and $\gamma_2 = 30$, respectively, indicating the transitions between regimes. (d-f) High-speed imaging snapshots of the suspension at $\dot{\gamma} = 1\,s^{-1}$, showing particle-scale configurations representative of the three regimes across the gap between the parallel plates.

sedimentation–resuspension transition, we performed strain-amplitude sweeps at varying angular frequencies. As shown in Figure 1(e), the critical strain thresholds required to initiate and sustain resuspension remain essentially constant across the experimentally accessible frequency range (1–20 rad/s). At very low angular frequencies, resuspension does not occur even at high strain amplitudes, likely because the Shields number remains below unity.

Our observations support this: at small strains, particle motion is limited to minor oscillations that do not disrupt gravitational settling. At large strains, chaotic particle collisions and rolling disrupt sediment structures, promoting upward transport and sustained dispersion. Thus, these findings instead point to a geometry-driven resuspension process in which particles must be displaced by a critical distance to initiate interactions and upward transport. The implications of this strain-governed framework will be further examined in various concentrations and flow modes in subsequent sections.

### B. Transient Dynamics of Particle Resuspension and Microstructural Evolution under Steady Shear Flow

To further probe the mechanisms behind particle resuspension, we performed time-resolved rheological measurements during the startup of steady shear flow, coupled with high-speed imaging to capture evolving microstructures. These experiments were designed to directly link bulk viscosity dynamics with particle-scale processes such as detachment, clustering, and orientation under shear. Understanding this transient behavior is critical in sediment-laden systems where gravitational confinement must be overcome by hydrodynamic forcing at low Shields number, conditions relevant to both geophysical flows and industrial suspensions.

We studied dense suspensions of PE particles dispersed in a Newtonian silicone oil, at a bulk volume fraction $\phi = 0.55$. After resting for 300 seconds to allow gravitational settling, the samples were subjected to constant shear rates of $0.2\,s^{-1}$, $1\,s^{-1}$, and $5\,s^{-1}$ using a parallel-plate rheometer. Simultaneously, high-speed imaging was used to visualize particle configurations across the flow gap. These paired measurements enabled us to characterize the temporal evolution of viscosity and corresponding particle dynamics under increasing strain.

As shown in Figure 2(a-c), the viscosity evolved in three distinct regimes, following the onset of shear. In Regime 1, we observed a rapid rise in viscosity immediately after shear is initiated. This phase corresponds to particle detachment from the sediment bed and the onset of individual motion into the fluid layer. In Regime 2, the viscosity continues to increase more gradually, indicating enhanced interparticle interactions and the formation of transient clusters and networks. Regime 3 is characterized by a plateau in viscosity, signifying a steady-state flow where the suspension has reached a dynamically resuspended and stabilized structure. Importantly, the durations of each regime varied with shear rate. At lower shear rates (e.g., $0.2\,s^{-1}$), Regime 2 persisted longer, reflecting a more gradual evolution of clustering and network formation. At the highest shear rate ($5\,s^{-1}$), Regime 2 was effectively bypassed, and the system transitioned rapidly from initial detachment to steady-state resuspension. This observation suggests that it is not shear rate alone, but the accumulated strain $\gamma$, that controls transitions between regimes. Across all cases, we identified two characteristic strain values: $\gamma_1 = 4$ and $\gamma_2 = 30$, marking the end of Regimes 1 and 2, respectively. These values appear robust across shear rates, indicating that particle-scale rearrangements follow a strain-governed pathway similar to the transitions observed in oscillatory shear, see Figure 1. At higher imposed shear rates, the duration of the intermediate stage between $\gamma_1$ and $\gamma_2$ becomes too short to be resolved, causing the two-step transition to merge into a single smooth rise in viscosity. For lower concentrations ($\phi \leq 0.50$), interparticle friction and packing constraints are weaker, leading to a single resuspension threshold. Moreover, near-wall layering observed in Regime 3 (See Video S2 (c) in Supplementary Material) reflects the intrinsic shear localization that accompanies yielding near the jamming limit. These effects do not alter the identification of the bulk resuspension thresholds, which was verified independently from both rheological and imaging data.

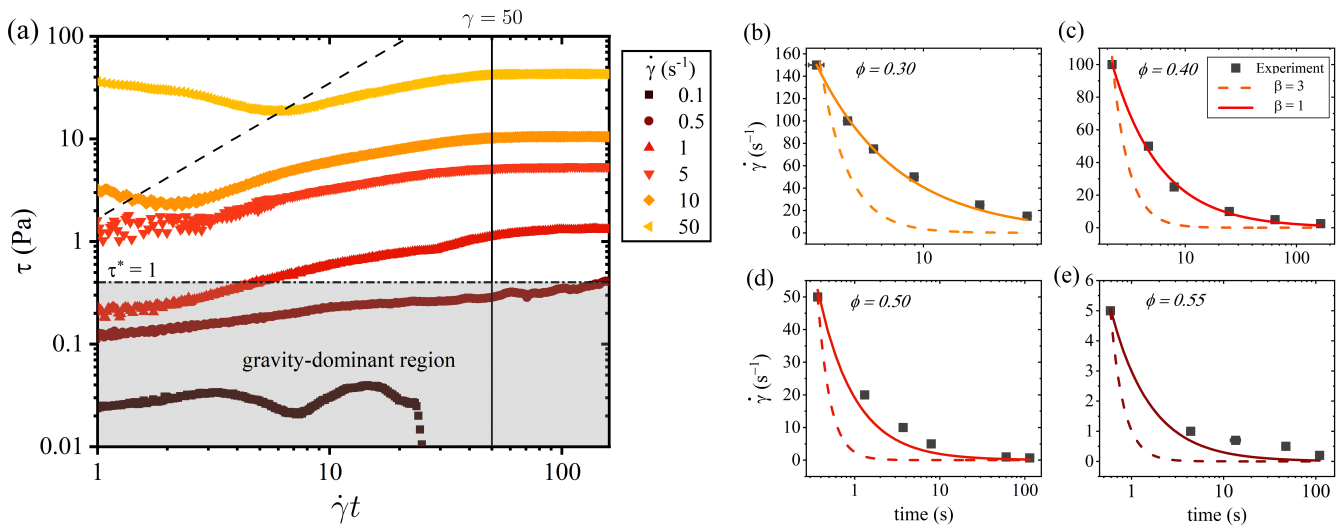

FIG. 3. Transient stress evolution and strain-governed resuspension dynamics under steady shear. (a) Shear stress plotted as a function of strain for a dense suspension with $\phi = 0.50$ under various shear rates. The dashed-dotted line and the shaded region highlight the gravity-dominated regime, defined by a dimensionless Shields number ($\tau^* < 1$). The solid vertical line at $\gamma = 50$ marks the critical strain at which all shear rates converge to a steady-state stress, suggesting a strain-controlled resuspension process. The dashed diagonal line shows the rheometer sensitivity limit. Applied shear rate as a function of time to reach steady state for various volume fractions : (b) $\phi = 0.30$, (c) $\phi = 0.40$, (d) $\phi = 0.50$, and (e) $\phi = 0.55$. The experimental data are represented by symbols, whereas the dashed and solid lines correspond to the fits of $\dot{\gamma} = k\Delta t^{-\beta}$ for the Acrivos[3] model ($\beta = 3$) and our proposed correlation ($\beta = 1$), respectively. The values of the error bars are smaller than the symbol size and are thus not shown here.

Direct visualizations from the 1 $s^{-1}$ case further elucidate these regimes. As shown in Figure 2(d), Regime 1 featured sparse particle motion confined to the fluid layer above the sediment bed. Few interparticle collisions occurred at this stage, but even limited detachment caused a sharp increase in viscosity. In Regime 2 (Figure 2(e)), collisions became more frequent, giving rise to isolated clusters and regions of collective motion. These structures correspond to intermediate states where hydrodynamic forcing begins to disrupt gravitational contacts. By Regime 3 (Figure 2(f)), the suspension exhibited coherent particle chains moving in aligned, horizontal trajectories, a signature of frictional flow and long-lived particle contacts. This regime represents a fully resuspended state under steady shear. An inverse correlation was observed between the extent of particle contacts and the rate of viscosity increase across these regimes. As particles reorganized into denser structures, additional strain yielded diminishing viscosity growth, indicating that frictional contacts dominate energy dissipation. This nonlinear rheological behavior is consistent with prior observations of frictional jamming and shear thickening in dense suspensions[25,43,44]. In the final steady-state regime (regime 3), the particles exhibit chain-like arrangements at the upper layer of the flow geometry, moving cohesively in a horizontal direction, shown in Figure 2(f). This behavior is characterized by continuous frictional contact among particles, a defining feature of this regime. An inverse relationship was observed between the extent of particle-particle contacts and the rate of viscosity increase, underscoring the nuanced dynamics of particle interactions.

Figure 2 (d-f) corresponding to each regime provide further insight into these dynamic processes, visualizing the onset of detachment, growth of collisional networks, and the eventual emergence of chain-like structures (Additional flow visualizations are provided in Video S2 (a-c) in Supplementary Material). Together, these results establish a mechanistic framework for understanding shear-induced resuspension under laminar, gravity-influenced conditions. By connecting transient rheological behavior with particle-scale organization, we identify strain, not just shear rate, as the key control parameter in determining the microstructural pathways toward resuspension in dense suspensions. Detailed observations of microstructural changes across flow regimes form the basis for subsequent analyses of particle clustering and collective behavior in complex flows.

### C. Shear Start-Up Behavior and Stress Evolution During Resuspension

To further understand the mechanisms governing shear-induced resuspension in dense suspensions, we investigated the time-resolved stress response during shear startup, with a focus on how these responses reflect evolving particle structures en route to steady state. These experiments complement the oscillatory measurements described in the previous section by directly probing how sedimented suspensions transition to resuspended flow when continuously sheared. Figure 3(a) presents the stress behavior, focusing on the representative case of $\phi = 0.50$ under various shear rates. Each sample was presheared at $\tau = 10$ Pa and allowed to rest for 300 s under gravity before initiating steady shear. At a low shear rate of 0.1 $s^{-1}$, the stress remains low and decreases over time, indicating that the imposed shear is insufficient to overcome gravitational forces and particle-particle friction, inhibiting the formation of an upward particle flux. After approximately 200 s, the shear stress rapidly decreases, as sedimentation causes the upper plate to come into contact with the pure fluid. The shaded area in Figure 3(a) marks this gravity-dominated regime (Shields number $\tau^* < 1$), where particle motion is primarily controlled by density contrast. At higher shear rates (1 $s^{-1}$ - 50 $s^{-1}$), the applied stress exceeds this threshold, and the system consistently reaches a steady stress value. We define the time to reach this steady state as the resuspension time, as it marks the point at which the volume fraction gradient dissipates and a fully dispersed suspension is formed. We define the resuspension time as the time to reach 95% of the stress plateau. Importantly, the data reveal that steady-state resuspension is consistently achieved once the strain reaches a critical value of $\gamma = 50$, regardless of the applied shear rate, marked by a solid vertical line in Figure 3(a). This observation further confirms that strain is a critical control parameter in the resuspension process, as it encapsulates both the viscous drag timescale for suspended particles, $t$, and the shear timescale associated with particle rearrangements, $\dot{\gamma}$. Although the final resuspended configuration represents a steady balance between the downward gravitational settling flux and the upward viscous resuspension flux, the transient evolution toward that state is governed primarily by cumulative strain rather than instantaneous shear rate. At small applied strains, particle motion is minimal and gravity dominates; as strain accumulates, progressive particle detachment and redistribution occur until the upward hydrodynamic flux balances gravity and a fully resuspended steady state is achieved. This interpretation reconciles the strain-controlled framework proposed here with the classical flux-balance picture, clarifying that strain history dictates the pathway to equilibrium, even though the final steady state satisfies a flux balance.

To interpret these results in light of existing theory, we compared our results with the model proposed by Acrivos et al.[3], which balances gravitational settling against shear-induced diffusive fluxes. According to this theory, the resuspension time ($\Delta t$) scales inversely with shear rate following:

$$\Delta t \propto \frac{A^{2/3} h_0^2}{\dot{\gamma} a^2}, \qquad (3)$$

where $h_0$ denotes the height of the sediment bed in the absence of shear, $a$ signifies the radius of the particles, and $A$ is a dimensionless parameter representing the ratio of the viscous force to the buoyant force, defined by $A = \frac{9}{2} \frac{\mu_0 \dot{\gamma}}{g \Delta \rho h_0}$, where $\mu_0$ and $\Delta \rho$ represent the liquid viscosity and particle-liquid density difference, respectively. When reformulated, the model predicts a power-law relationship between shear rate and resuspension time:

$$\dot{\gamma} = k \Delta t^{-\beta}, \qquad (4)$$

where, $k$ represents a constant and $\beta = 3$ based on asymptotic assumptions.

We fit this expression to our experimental data for each particle concentration, defined as the time to reach 95% of the steady-state stress. As shown in Figures 3(b–e) we find that the best fit is obtained with $\beta = 1$ across all four volume fractions ($\phi = 0.30, 0.4, 0.50, 0.55$), not $\beta = 3$. This result indicates that the dominant control parameter is strain, not shear rate, suggesting that the cumulative deformation history dictates the resuspension dynamics in dense suspensions.

This distinction is crucial. The Acrivos model assumes that resuspension is dominated by the competition between steady fluxes: gravitational settling and shear-induced diffusion. However, our findings suggest that such steady-state assumptions are insufficient for dense suspensions, particularly at higher particle concentrations where transient microstructure, interparticle friction, and contact-dominated rearrangements dominate the dynamics. These factors introduce a path-dependent rheology and delay resuspension beyond what is predicted by purely diffusive models. Under this framework, achieving a uniform suspension corresponds to reaching a critical total strain, $\gamma = k$, which is consistent across varying shear conditions but depends on the volume fraction.

The alignment of the experimental curves with this theoretical framework when $\beta = 1$ suggests that strain, rather than the shear rate, is the more relevant parameter controlling the capability of the system to reach a steady state. However, the model proposed by[3], in which $\beta$ was set to 3, exhibits a consistent discrepancy when compared quantitatively to the experimental data, particularly at higher particle concentrations and lower shear rates. These deviations suggest that the model's asymptotic assumptions, such as neglecting transient microstructural evolution, finite particle interactions, and history-dependent rheology, maybe overly simplistic for describing dense suspensions. In such regimes, collective particle rearrangements, short-range hydrodynamic interactions, and local jamming dynamics become nonnegligible and are not captured by the continuum-level approximations of the Acrivos framework[3]. This discrepancy is not merely a quantitative mismatch; it points to a qualitative limitation in how the mechanisms of shear-induced resuspension are conceptualized in current theoretical models. The experimental results reveal that the process by which the system reaches a steady state is not governed solely by the instantaneous shear rate but also by the strain, a parameter that more directly accounts for the cumulative deformation history and associated microstructural transformations within the suspension. These insights, drawn from experiments conducted under steady shear conditions, align with those obtained under oscillatory shear conditions, where the strain amplitude governs the transition from a sedimented to a resuspended state.

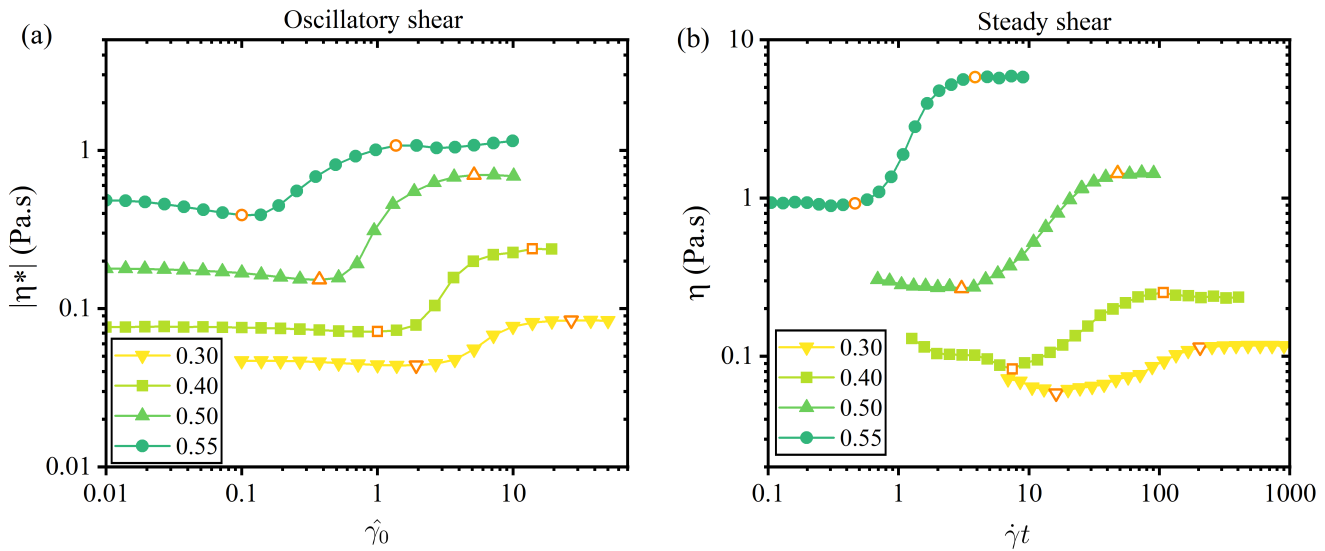

FIG. 4. Comparison of resuspension behavior under oscillatory and steady shear across different particle volume fractions ranging from 0.30 to 0.55. (a) Magnitude of complex viscosity $|\eta^*|$ as function of strain amplitude at a fixed angular frequency of $\omega = 10$ rad/s.(b) Viscosity $\eta$ as a function of strain under steady shear conditions, measured from flow ramp experiments on the same suspensions. Both panels show two distinct transitions: the onset of resuspension ($\gamma_i$) and its completion ($\gamma_c$), marked by open symbols.

Collectively, these results reveal that strain, whether accumulated over time (steady shear) or applied cyclically as an amplitude (oscillatory shear), is a unifying control parameter for particle resuspension in dense suspensions. We denote this dimensionless strain as $\gamma$ in steady shear and $\hat{\gamma}$ in oscillatory shear throughout the paper. The strain reported here refers to the macroscopic engineering strain applied by the rheometer, defined as the top-plate displacement divided by the fixed gap height. In partially sedimented suspensions, this global deformation is initially borne almost entirely by the interstitial fluid above the packed bed. As the applied strain increases, hydrodynamic stresses are transmitted downward, progressively shearing and mobilizing the particle packing.

The Acrivos–Fan scaling (Eq. (3)) was derived for a narrow-gap Couette geometry under steady, uniform shear, whereas the present configuration involves time-dependent stress and nonuniform shear across the gap. Nevertheless, Acrivos et al.,[3] showed that the shear-induced diffusivities along and normal to the shear plane ($D_\parallel$ and $D_\perp$) are nearly equivalent under constant shear, supporting the applicability of the same diffusive framework to parallel-plate systems. Thus, the correlations among resuspension height, stress, and time remain qualitatively valid as long as the Shields number is moderate and the flow is quasi-steady. In our experiments, however, both $\tau$ and $Sh$ evolve continuously during transient resuspension, making the problem strongly nonlinear and requiring numerical treatment for full resolution. Our results therefore extend these classical models to dense, gravity-dominated suspensions where cumulative strain—not time—governs the resuspension dynamics. This conclusion is supported by our prior numerical studies[45–47], which resolved spatiotemporal concentration and stress variations and confirmed that strain, rather than time, controls the onset and completion of resuspension, consistent with the scaling derived by Chapman & Leighton Jr.[48] where transient strain scales as $(h/a)^2$.

## IV. DISCUSSION

To systematically investigate the role of the bulk particle concentration on resuspension dynamics, we conducted a series of oscillatory shear experiments in which the magnitude of the complex viscosity, $|\eta^*|$, was measured as a function of increasing strain amplitude for suspensions with particle volume fractions ranging from $\phi = 0.30$ to $\phi = 0.55$, as shown in Figure 4(a). The applied shear strain remains insufficient to overcome particle-particle friction and gravitational settling at small strain amplitudes. Consequently, particle motion is negligible, and the suspension remains in a sedimented state, reflected by a plateau in the magnitude of the complex viscosity. However, as the strain amplitude increases, $|\eta^*|$ exhibits a sharp rise, indicating the onset of particle collisions and resuspension. This transition arises from the initiation of interparticle collisions and detachment events as the sedimented particles begin to move, owing to the inertial and viscous forces of the fluid. In particular, the critical strain amplitude required to trigger resuspension decreases as the particle volume fraction increases. Denser suspensions exhibit transitions to resuspension at smaller strain amplitudes $\hat{\gamma}_0$, which is reflective of an increased sensitivity to deformation due to reduced interparticle spacing and increased collective stress transmission. This observation reinforces previous reports that particle migration in non-Brownian suspensions is predominantly governed by the magnitude of the applied strain amplitude, highlighting the critical role of strain-controlled dynamics in such systems[49–53].

To extend these observations to steady shear conditions and develop a unified framework for analyzing resuspension across various shear modes, we performed strain-controlled flow ramp experiments. We show in Figure 4(b), the viscosity $\eta$ as a function of the strain for each volume fraction. Although the oscillatory and steady shear modes fundamentally differ in terms of temporal characteristics, a similar characteristic of the resuspension process is observed under both conditions: (1) a sharp increase marking the onset of resuspension ($\gamma_i$), and a plateau corresponding to a fully suspended state ($\gamma_c$) which indicates the consistency of the strain-controlled resuspension under steady and dynamic shear modes.

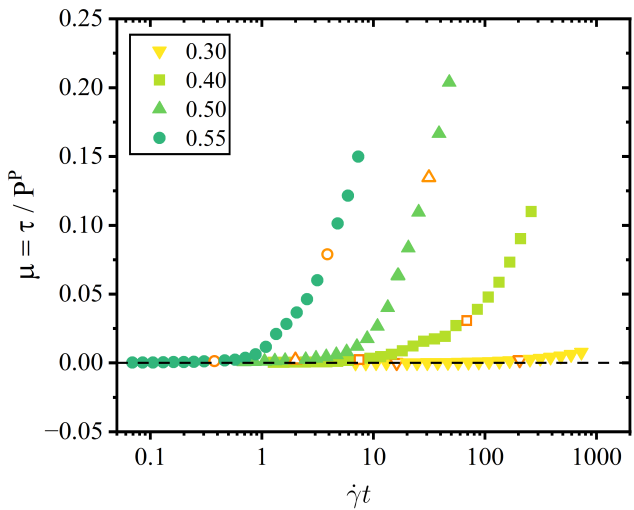

FIG. 5. Evolution of the effective friction coefficient $\mu(\gamma) = \tau/P^P$ as a function of strain for suspensions with different particle volume fractions $\phi = 0.30, 0.40, 0.50,$ and $0.55$. The dashed horizontal line at $\mu = 0$ corresponds to the initial sedimented state with negligible interparticle contact or normal stress transmission. Open symbols denote the critical strains marking the initiation ($\gamma_i$) and completion ($\gamma_c$) of the resuspension process.

To probe the mechanical origin of these transitions, we measured the effective friction coefficient $\mu(\gamma)$, under steady shear conditions. The effective friction coefficient is defined as the ratio of the applied shear stress to particle-confining pressure[54]: $\mu(\gamma) = \frac{\tau}{P^P}$, a dimensionless parameter that reflects the balance between the tangential and normal forces within the suspension.

The particle-confining pressure, $P^P$, was determined experimentally from the normal force exerted on the rheometer plates, normalized by the cross-sectional area. As shown in Figure 5, $\mu(\gamma) \approx 0$, confirming the absence of normal stress and interparticle contact. Once $\gamma$ exceeds the initiation strain $\gamma_i$, $\mu(\gamma)$ increases sharply, marking the onset of shear-induced particle interactions, including collisions and contact friction. The slope and magnitude of this grow with $\phi$, consistent with predictions from frictional rheology and jamming theory[54,55]. In these frameworks, such behaviors are attributed to enhanced interparticle constraints as the packing density increases. These observations reinforce the idea that particle resuspension is governed not solely by fluid-induced forces but also by strain-mediated frictional mechanisms that depend sensitively on the system's microstructure.

For strain amplitudes exceeding $\gamma_c$, the continued rise of the effective friction coefficient $\mu(\gamma)$ reflects the evolving balance between tangential and normal stresses once the suspension is fully mobilized. In this post-resuspension regime, the particle pressure $P^P$ remains nearly constant, as the particle phase is homogeneously distributed. However, under the flow-ramp protocol, the shear rate—and consequently the shear stress $\tau$—continues to increase with strain, producing a gradual rise in $\mu = \tau/P^P$. This behavior indicates a transition from a quasi-static, friction-dominated state to a hydrodynamic–collisional regime, where stress transmission is governed by lubrication forces and transient contacts. Similar post-yield increases in $\mu$ have been reported in dense suspensions near the jamming limit[18,54], where residual microstructural anisotropy and finite contact networks sustain normal stresses even after macroscopic homogeneity is achieved.

To synthesize these insights into a unified framework, we constructed state diagrams that delineate the boundaries between sedimented, partially resuspended, and fully suspended states in the $(\phi, \gamma)$ plane under both oscillatory and steady shear (Figure 6(a,b)). These diagrams are derived directly from the experimentally identified thresholds $\gamma_i$ and $\gamma_c$ obtained in both deformation modes. Both thresholds decrease systematically with increasing $\phi$, consistent with the reduced mean particle spacing and the enhanced frequency of frictional contacts at higher packing densities. As the interstitial fluid fraction diminishes, particle–particle collisions become more frequent, promoting collective motion and bulk mobilization at smaller imposed deformations. A comparison between the two maps reveals that the strains required for resuspension under steady shear are nearly an order of magnitude larger than those in oscillatory shear. This contrast highlights the role of oscillatory motion in enhancing hydrodynamic interactions and particle collisions, thereby facilitating bed fluidization through reversible–irreversible transitions in particle trajectories. In particular, the critical strain for resuspension scales robustly with particle concentration, confirming that accumulated deformation, rather than instantaneous shear rate, governs the resuspension process.

It should also be noted that both the cumulative strain and the bulk particle volume fraction plotted in Figure 6 are macroscopic, geometry-averaged quantities rather than intrinsic local variables. In the early stages of deformation, when a portion of the suspension remains sedimented, the shear field is highly heterogeneous across the gap. Nevertheless, the apparent cumulative strain provides a robust global measure of the total deformation required to mobilize the bed. Similarly, the bulk volume fraction, though an averaged quantity, effectively parametrizes the solid loading that governs the stress balance and resuspension threshold. Consequently, the phase diagrams in Figure 6 should be regarded as a phenomenological representation of the global system state, constructed from experimentally accessible quantities.

To quantitatively capture the concentration dependence of the critical strain, we propose a scaling relation based on the mean interparticle separation:

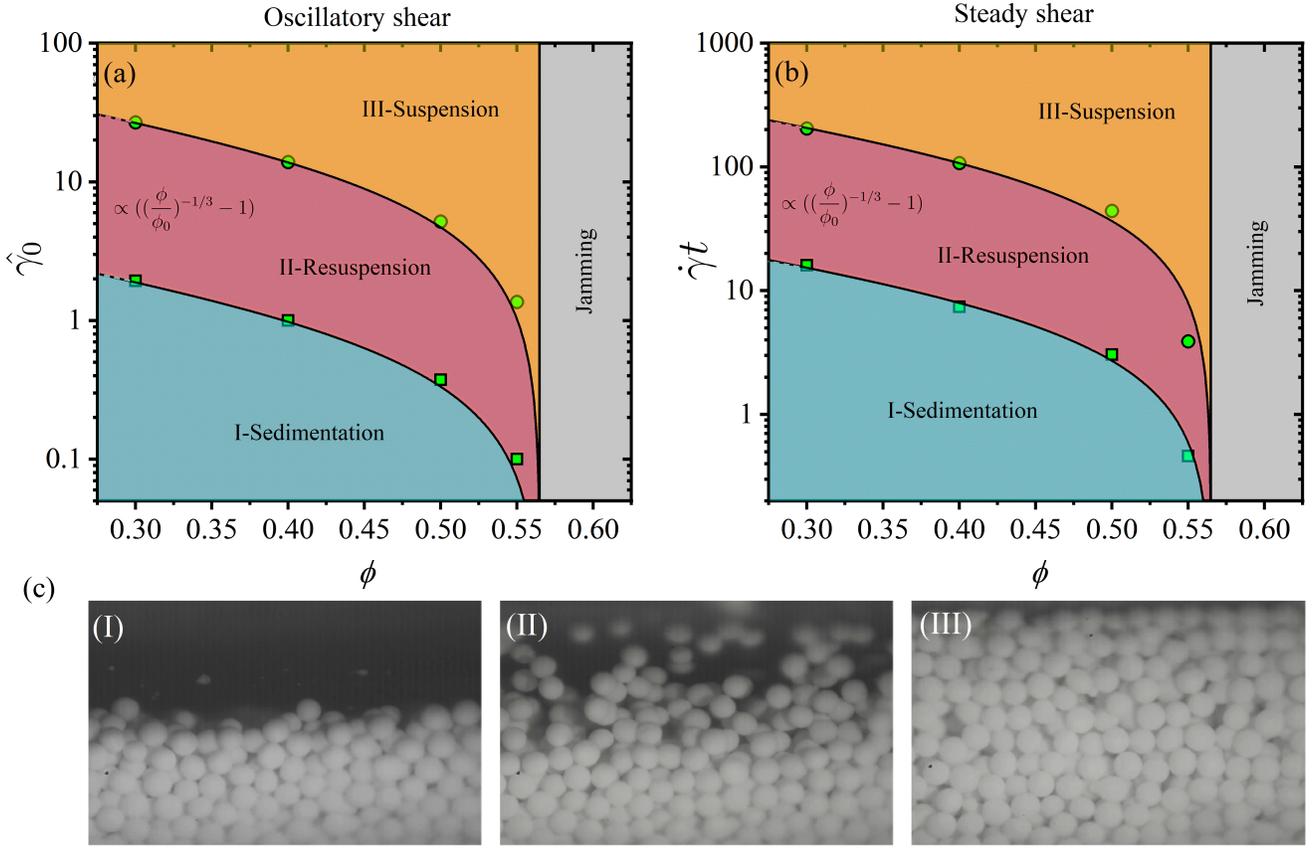

FIG. 6. State diagrams and microstructural snapshots in dense suspensions. (a, b) Regime maps for oscillatory and steady shear flows, respectively, were constructed over a wide range of particle volume fractions ($\phi = 0.30$ to $0.55$). The diagrams delineate four distinct flow regimes: sedimentation (green), resuspension (purple), fully suspended flow (orange), and jamming (gray). Squares and circles represent the experimentally determined critical strains for the initiation ($\gamma_i$) and completion ($\gamma_c$) of resuspension for each volume fraction. Solid black curves correspond to the predictive fit based on Eq. (5). Error bars are smaller than the symbol size and are not shown for clarity. (c) Representative side-view images from the rheometer gap show particle configurations in sedimented (I), resuspended (II), and fully suspended (III) states.

$$\gamma = k \left( \left( \frac{\phi}{\phi_0} \right)^{-1/3} - 1 \right). \quad (5)$$

where $k$ is a fitting parameter and $\phi_0$ is the maximum particle packing fraction, which is approximately 0.57 for non-Brownian spheres, as reported in[43,54,55]. A detailed derivation of this expression is provided in the Supplementary Information. This scaling captures the idea that the strain required for particles to make contact—and thus initiate frictional rearrangement—decreases as the average center-to-center distance diminishes with increasing $\phi$. While this formulation is not a first-principles model, it serves as a semi-empirical predictive framework grounded in the physical scaling of particle–particle collisions. The dimensionless prefactor $k$ is determined once for each flow type (oscillatory or steady) and then applied across all tested particle concentrations without further adjustment. In this sense, the model is calibrated but predictive in its application, capturing the dependence of resuspension thresholds on both concentration and deformation mode through a single, physically interpretable relation.

This distinction clarifies that our framework provides predictive capability within its empirical range, rather than a purely theoretical derivation. In addition, it should be emphasized that Eq. (5) represents a geometry-averaged, phenomenological scaling between macroscopic quantities rather than a local constitutive law, intended to capture the global strain required for collective bed mobilization across heterogeneous concentration fields.

To visualize the evolution of particle microstructure during resuspension, particle-scale images were captured across the rheometer gap. As shown in Figure 6c, the system evolves through three characteristic regimes: (I) an initial sedimented state where particles rest at the bottom, (II) a resuspension regime marked by partial particle detachment and the formation of a concentration gradient, and (III) a fully suspended regime in which particles are uniformly dispersed throughout the gap. This sequence demonstrates that resuspension is not an instantaneous process but proceeds through intermediate stages in which particle interactions, clustering, and vertical concentration gradients evolve continuously with increasing strain.

Although the Shields number and the strain amplitude are not fully independent, they describe complementary aspects of the resuspension process. The Shields number quantifies the instantaneous force balance between hydrodynamic and gravitational stresses, determining whether the local stress magnitude is sufficient for particle detachment. In contrast, the strain amplitude governs the cumulative deformation required to progressively disrupt the contact and frictional networks within the sedimented bed. In the regime where $Sh \approx 1$, that is, when the shear and gravitational stresses are comparable, the onset and completion of resuspension are primarily dictated by the total strain accumulated over time. This explains why the resuspension thresholds identified in Figures 1(b) and 3(a) depend more strongly on strain than on instantaneous shear rate. Thus, $Sh$ establishes the necessary stress level for resuspension to occur, while the strain determines the degree to which the suspension has been deformed sufficiently to achieve and maintain a fully resuspended steady state.

## V. CONCLUSIONS

In this study, we demonstrated that resuspension of non-Brownian particles in Newtonian fluids is fundamentally governed by strain, rather than shear rate alone—a departure from conventional frameworks that have largely emphasized rate-based control. We investigate the resuspension dynamics of non-Brownian spherical particles dispersed in a Newtonian fluid using rheological measurements conducted under both oscillatory and steady shear conditions. By analyzing suspensions across a broad range of particle volume fractions ($\phi$ = 0.30 to 0.55), we showed that the process of resuspension is governed not by the shear rate alone but by the strain, with distinct critical strain ranges for the onset and completion of resuspension observed for each volume fraction. We have shown that these strain thresholds are largely independent of the rest time.

Our rheological measurements revealed that the initiation of resuspension is marked by a sharp increase in the magnitude of the complex viscosity, which is attributed to the onset of effective particle-particle collisions. These interactions result in the generation of hydrodynamic and frictional forces that are sufficient to overcome sedimentation and gravitational constraints, promoting particle lift-off and redistribution within the fluid. Furthermore, through experiments with increasing stress, we demonstrated that the resuspension time scales inversely with the applied shear rate as $\dot{\gamma} \propto t^{-1}$, supporting our major finding that sufficient strain must be acquired to initiate resuspension.

To generalize our findings, we derived a scaling relation that connects the critical strain for resuspension to the particle volume fraction. On the basis of this formulation, we constructed state diagrams that clearly delineate the transition between the sedimented, resuspended, and fully suspended regimes for both the oscillatory and steady shear modes. Our phase diagrams reveal a significant contrast between flow modalities: the resuspension strain thresholds under steady shear conditions are nearly an order of magnitude greater than those under oscillatory shear conditions, highlighting the increased efficiency of oscillatory flows in inducing collisions and mobilizing dense suspensions.

By integrating rheometric observations with in situ optical imaging, we identified two key mechanisms underpinning the resuspension process: (i) strain-driven effective collisions that initiate particle detachment and (ii) collective motion and rearrangement that accelerate particle detachment and sustain suspension. Together, these findings provide a comprehensive framework for understanding and predicting particle resuspension dynamics in dense, non-Brownian suspensions. This new understanding is expected to have broad implications for processes in which control over particle dispersions is critical, such as industrial processes, environmental transport, and the design of functional materials.

## VI. SUPPLEMENTARY MATERIAL

The supplementary material includes additional figures, Videos, and detailed calculations that further support the results presented in this study.

## ACKNOWLEDGMENTS

The authors are grateful to Dr. Fredric Blanc for his insightful discussions and valuable feedback. P.M. and M.M. acknowledge support in part from the National Science Foundation under Grant Nos. NSF-CBET-2230892 and NSF-CBET-2335195.

## DATA AVAILABILITY STATEMENT

The data that support the findings of this study are available from the corresponding author upon reasonable request. All relevant data generated or analyzed during this study are included in this published article and its supplementary information files.

packed suspensions of different particle shapes," Physical Review E **84**, 031408 (2011).